\documentclass[a4paper,12pt]{llncs}
\usepackage{amsmath,amssymb}
\usepackage{hyperref}
\usepackage[left=3.2cm,top=3.2cm,right=3.2cm,bottom=3.2cm]{geometry}
\usepackage{epsfig}

\usepackage{color}
\usepackage{graphicx}
\usepackage{subfigure}
\usepackage{floatrow}
\newfloatcommand{capbtabbox}{table}[][\FBwidth]

\usepackage{wrapfig}
\usepackage{amssymb}
\usepackage{amsmath}
\usepackage{tikz, hyperref}

\usepackage{longtable}
\usepackage{multirow}
\usepackage{array}
\usepackage{paralist}

\date{}
\begin{document}
\title{\Large{Dynamic Competition Networks: detecting alliances and leaders}\thanks{Research supported by grants from NSERC, Ryerson University, NSF CCF-1149756, IIS-1546488, NSF Center for Science of Information, CCF-0939370, DARPA SIMPLEX.}}
\author{Anthony Bonato\inst{1}, Nicole Eikmeier\inst{2}, David F. Gleich\inst{2}, \and Rehan Malik \inst{1}}
\institute{Ryerson University \and Purdue University}
\maketitle
\begin{abstract}

We consider social networks of competing agents that evolve dynamically over time. Such dynamic competition networks are directed, where a directed edge from nodes $u$ to $v$ corresponds a negative social interaction.  We present a novel hypothesis that serves as a predictive tool to uncover alliances and leaders within dynamic competition networks. Our focus is in the present study is to validate it on competitive networks arising from social game shows such as Survivor and Big Brother.

\end{abstract}

\section{Introduction}\label{intro}

Complex social networks are heterogeneous, evolving, and pervasive in the natural world and in technological settings. Social networks present rich sources of complex networks, where nodes represent
agents and edges correspond to some form of social interaction. For example, in Facebook edges represent friendship, while on Twitter they denote following. Complex, social networks commonly display
power law degree distributions, the small world property (short distances between nodes and high local clustering) and other phenomena such as densification and strong community structure; see
\cite{bbook,at,k}. Another key principle underlying social networks is that links exhibit homophily, that is, nodes with similar social attributes are linked, which is related to an embedding of the nodes in a so-called \emph{Blau space}, where nodes are assigned to points in a suitable metric space and the relative distance
between pairs of nodes is a function of similar social attributes. See \cite{BG,MR}.

While social interaction is usually studied from the premise of friendship, cooperation, or other positive social interactions, there is a growing literature on the study of \emph{negative}
social interaction as a generative mechanism underlying social networks. For example, while transitivity is a folkloric notion in social networks, summarized in the adage that ``friends of friends
are more likely friends,'' structural balance theory (see \cite{he} and \cite{k} for a modern treatment) points also to the inverse adage ``enemies of enemies are more likely friends.'' A common problem in this direction is the prediction of the type of edges in a social network~\cite{predict,negative,frenemy}. Hence,
competitive and negative relationships are critically important to the study of social networks, and are often hidden drivers of link formation.

Competitive relationships were studied recently via the Iterated Local Anti-Transitivity (or ILAT) model; see \cite{ilt1,ilt}. In the ILAT model, each node $u$ duplicates every time-step by forming
its \emph{anti-clone} $u'$, so that $u'$ joins to the nodes in the non-neighbor set of $u$. We may also consider real-world networks of opposing nation states, rival gangs or other organizations, and
consider alliances formed by mutually shared adversaries. The ILAT model provably generates highly dense graphs with low diameter and high local clustering. See \cite{guo} for a recent study using
the spatial location of cities to form an interaction network, where links enable the flow of cultural influence, and may be used to predict the rise of conflicts and violence. Another example comes
from market graphs, where the nodes are stocks, and stocks are adjacent as a function of their correlation measured by a threshold value $\theta \in (0,1).$ Market graphs were considered in the case
of negatively correlated (or competitive) stocks, where stocks are adjacent if $\theta < \alpha,$ for some positive $\alpha$; see \cite{market}.

In the present paper, we focus on the underlying structure of social networks of competitors that evolve dynamically over time. We view such networks as directed, where a directed edge
from nodes $u$ to $v$ corresponds to some kind of negative social interaction. For example, a directed edge may represent a vote by one player for another in a social game such as the television
program Survivor. Directed edges are added over discrete time-steps in what we call dynamic competitive networks. Our main contribution in this empirical work is a hypothesis that serves as a
predictive tool to uncover alliances and leaders within dynamic competition networks. While the hypothesis may hold more broadly, our focus here is on competitive networks arising from social game shows. We validate the hypothesis using voting record data of the social game shows Survivor and Big Brother.

We organize the discussion in this paper as follows. In Section~\ref{model}, we formally introduce dynamic competition networks, and using graph theoretic tools, give a precise
formulation of the Dynamic Competition Hypothesis. In Section~\ref{distance} and the Appendix (which will appear in the full version of the paper), we present voting data from all the seasons of U.S.\ Survivor and Big Brother, focusing on three seasons of Survivor in detail and one season of Big Brother. We analyze this data using tools
from network science in an effort to validate the Dynamic Competition Hypothesis. We find that the hypothesis accurately predicts the emergence of alliances and predicts finalists with a high degree
of precision. The final section interprets our results within the context of real-world complex
networks, and presents open problems derived from our analysis.

We consider directed graphs with multiple directed edges throughout the paper. For background on graph theory, the reader is directed to~\cite{west}. Additional background on complex networks may be
found in the book \cite{bbook}.

\section{Dynamic Competition Hypothesis}\label{model}

 A \emph{competition network} $G$ is one where nodes represent agents, and there is directed edge between nodes $u$ and $v$ in $G$ if agent $u$ is in competition with agent $v$. The directed edge
$(u,v)$ may also represent a vote against $v$ (depending on the nature of $G$). A \emph{dynamic competition network} is a competition network where directed edges are added over discrete time-steps.
For example, on the game show Survivor (as we discuss in detail in the next section), players cast votes against each other, and the votes correspond to directed edges in the network. As another
example, nodes may consist of nation states and edges correspond to conflicts between them. Dynamic competition networks may have multiple edges. Note that dynamic competition networks are also
models of (sports) tournaments. However, in dynamic competition networks, not all nodes are joined by edges as is typically the case in tournaments. Our focus in this work will be on dynamic competition networks arising in social networks, and we focus specifically on networks arising from Survivor and Big Brother.

Before we describe our hypothesis about the structure of competition networks, we present some graph-theoretic terminology. We consider standard metrics in network science, such as in- and out-degree, closeness and betweenness. Given the nature of the voting network in Survivor, we also consider the number of common out-neighbors as a key metric.

For nodes $u,$ $v,$ and $w,$  we say that $w$ is a \emph{common out-neighbor} of $u$ and $v$ if $(u,w)$ and $(v,w)$ are directed edges. For a pair of distinct nodes $u,v$, we define
$\mathrm{CON}(u,v)$ to be the number of common out-neighbors of $u$ and $v$. For a fixed node $u$, define $$\mathrm{CON}(u)=\sum_{v\in V(G)} \mathrm{CON}(u,v).$$ We call $\mathrm{CON}(u)$ the
\emph{CON score} of $u.$  For a set of vertices $S$ with at least two nodes, we define $$\mathrm{CON}(S) = \sum_{u,v\in S} \mathrm{CON}(u,v).$$ Note that $\mathrm{CON}(S)$ is a non-negative integer.

A set of nodes $S$ with no directed edges in its induced subgraph is called \emph{independent}; we also need a notion of being ``close'' to independent. For a set $S$ of nodes, define its \emph{edge density} to be the ratio
$ED(S)=|E(S)|/\binom{|S|}{2}.$ Observe that $ED(S)$ may be greater than 1 as there may be multiple edges in the digraphs we consider. For a non-negative real number $\epsilon$ say that a set $S$ is
$\epsilon$-\emph{near independent} if $ED(S)\le \epsilon$. The parameter $\epsilon$ measures the relative density of sets of vertices. We say that a set is near independent if it is
$\epsilon$-\emph{near independent} for some positive value of $\epsilon$; typically, in applications, we take $\epsilon$ to be small. The value of $\epsilon$ will often be heuristically determined in
a real-world networks by considering a ranking of subsets by their edge density. Note that independent sets are trivially near independent.

For a strongly connected digraph $G$ and a node $v$, define the \emph{closeness} of $u$ by $$C(u)=\left(\sum _{v \in V(G)\setminus \{u \}}  d(u,v)\right)^{-1}$$  where $d(u,v)$ corresponds to the distance measured by one-way,
directed paths from $u$ to $v$. The \emph{betweenness} of $v$ is defined by $$B(v)=\sum _{x,y \in V(G)\setminus \{v \}} \sigma_{xy}(v)/\sigma_{xy},$$ where $\sigma_{xy}(v)$ is the number of shortest
one-way, directed paths between $x$ and $y$ that go through $v$, and $\sigma_{xy}$ is the number of shortest one-way, oriented paths between $x$ and $y$. Both closeness and betweenness are well-studied
centrality measures for complex networks~\cite{brandes}. For example, centrality of sports networks is often used to rank teams~\cite{whos}.

\subsection{The hypothesis}

\emph{Alliances} are defined as groups of agents who pool capital towards mutual goals. In the context of social game shows such as Survivor, alliances are groups of players who work together to vote off players outside the alliance. Members of an alliance are typically less likely to vote for each other, and this is the case in strong alliances. \emph{Leaders} are defined as members with high standing in the network, and edges emanating from leaders may
influence edge creation in other agents. In Survivor, leaders may be the winner of a given season, but may also be non-winning players with a strong influence on the outcomes of the game. One of our main goals is to apply network science to help determine alliances and leaders in dynamic competitive networks arising in social networks.

The \emph{Dynamic Competition Hypothesis} (or \emph{DCH}) asserts that dynamic competition networks arising from a social networks satisfy the following four properties.

\begin{enumerate}
\item Alliances are near independent sets.
\item Strong alliances have low edge density.
\item Members of an alliance with high CON scores are more likely leaders.
\item Leaders exhibit high closeness, high CON scores, low in-degree, and high out-degree.
\end{enumerate}

The DCH provides a quantitative framework for the structure of dynamic competition networks arising from social networks; no other data is required other than the presence of competitive relationships. See Figure~1
for a visualization of the DCH.

\begin{figure}
\begin{center}
\epsfig{figure=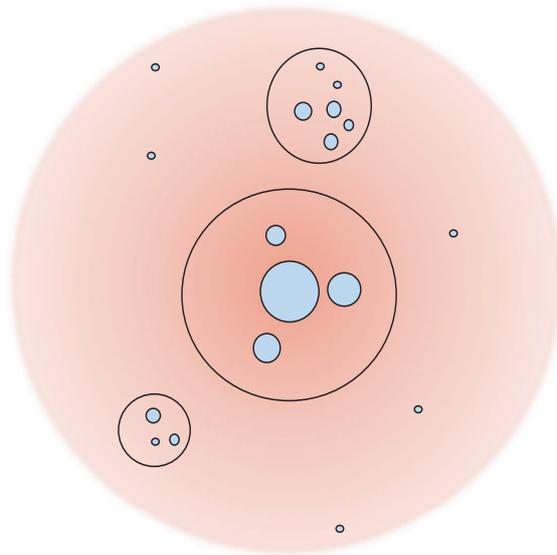,width=3in,height=3in}
\end{center}
\caption{A heat map representation of dynamic competition networks according to the DCH, where nodes closer to the center have higher closeness and CON scores. Larger nodes have higher CON scores, lower in-degree, and higher out-degree. The subsets correspond to alliances.}
\end{figure}

Note how items (1), (2), and (3) mutually reinforce each other. Once we have discovered an alliance as per (1), we can measure its strength relative to other alliances via (2), and use (3) as tool to
isolate leaders within alliances. Item (4) is independent of alliances; in particular, while we expect leaders to be in alliances (that is, have prominent local influence), leaders are determined via
global metrics of the network.

Interestingly, closeness rather than betweenness appears be a good centrality measure in the dynamic competition networks studied in the next section. This may be explained by the low in-degree of nodes corresponding to leaders.

\section{Data and Methods}\label{distance}

We extracted data from the American television series Survivor over all of its seasons, and for further validation, from all seasons of Big Brother. Before we present the data in detail for a subset of seasons, we give some background
on both series. Survivor and Big Brother are examples of social games, where social interactions help determine the gameplay and winner. We focus on the US version of both shows, but they play in
several countries, accounting for over one hundred seasons in total.

In Survivor, strangers called \emph{survivors} are placed in a location and forced to provide shelter and food for themselves, with limited support from the outside world. Survivors are split into
two or more \emph{tribes} which cohabitate and work together. Tribes compete for immunity and the losing tribe goes to tribal council where one of their members is voted off. At some point during the
season, tribes merge and the remaining survivors compete for individual immunity. Survivors voted off may be part of the \emph{jury}. When there are a small number of remaining survivors who are
\emph{finalists} (typically two or three), the jury votes in favor of one of them to become the \emph{Sole Survivor} who receives a cash prize of one million dollars.

In Big Brother, a group of strangers called \emph{HouseGuests} cohabitate in a custom set under video surveillance. Each week, the HouseGuests compete for the title of \emph{Head of Household}, who
must nominate two HouseGuests for eviction. The Houseguests vote to evict one of them, and the one with the most votes is evicted. The winner received a cash prize of half a million dollars.

In both Survivor and Big Brother, several twists have been introduced during the seasons. For example, in Survivor, these include the introduction of a hidden immunity idol which would protect a
survivor from being voted out if used during tribal council. As a disclaimer, our analysis is insensitive to these twists.

Data was taken from Survivor Wiki \cite{SW} and Big Brother Wiki \cite{BBW}, which contains information on contestants, their voting records and tribes, and catalogues of alliances. For computing
centrality metrics and for the dynamic competition graph visualization, we used the open source Gephi software \cite{g}.

We present below visualizations of the dynamic competition networks for Survivor: Borneo, China, Game Changes, and HHH; we also include data from Season~12 of Big Brother. Note that the data is taken
after all votes had been cast against other players, and tables are provided with a summary of relevant network statistics. The order of the tables is given by their elimination order from the game, so the first entry is the winner and the others are ordered by when they were eliminated. In all of the five seasons described below, the data conforms to the
predictions of the DCH with regards to leaders (that is, winners in this context). It also clearly delineates alliances, as we discuss below.

\subsection{Borneo}

We consider the first season of Survivor set in Borneo. The abbreviations ID, OD, C, CON, and B stand for in-degree, out-degree, closeness, CON-score, and betweenness, respectively.
\begin{center}
\begin{tabular}{|l|l|l|l|c|l|c|}
\hline

Name & Tribe & ID & OD & C & CON & B\\ \hline \hline

Richard & Tagi & 6 & 10 & 0.737 & 42 & 28.7\\ \hline
Kelly & Tagi & 0 & 12 & 0.682 & 34 & 0\\ \hline
Rudy & Tagi & 8 & 11 & 0.778 & 45 & 36.483\\ \hline
Susan & Tagi & 7 & 10 & 0.778 & 44 & 16.467\\ \hline
Sean & Tagi & 9 & 9 & 0.7 & 38 & 17.917\\ \hline
Colleen & Pagong & 7 & 8 & 0.636 & 29 & 33.067\\ \hline
Gervaise & Pagong & 6 & 7 & 0.636 & 31 & 8.583\\ \hline
Jenna & Pagong & 11 & 6 & 0.583 & 27 & 27.85 \\ \hline
Greg & Pagong & 6 & 5 & 0.412 & 15 & 4.833\\ \hline
Gretchen & Pagong & 4 & 4 & 0.56 & 17 & 7.233\\ \hline
Joel & Pagong & 4 & 3 & 0.412 & 17 & 1\\ \hline
Dirk & Tagi & 4 & 3 & 0.5 & 12 & 1.317\\ \hline
Ramona & Pagong & 6 & 2 & 0.412 & 10 & 17.733\\ \hline
Stacey & Tagi & 6 & 2 & 0.452 & 4 & 1.733\\ \hline
B.B. & Pagong & 6 & 1 & 0.298 & 5 & 0.333\\ \hline
Sonja & Tagi & 4 & 1 & 0.452 & 4 & 0.75\\ \hline
\end{tabular}
\begin{figure}[h!]
\epsfig{figure=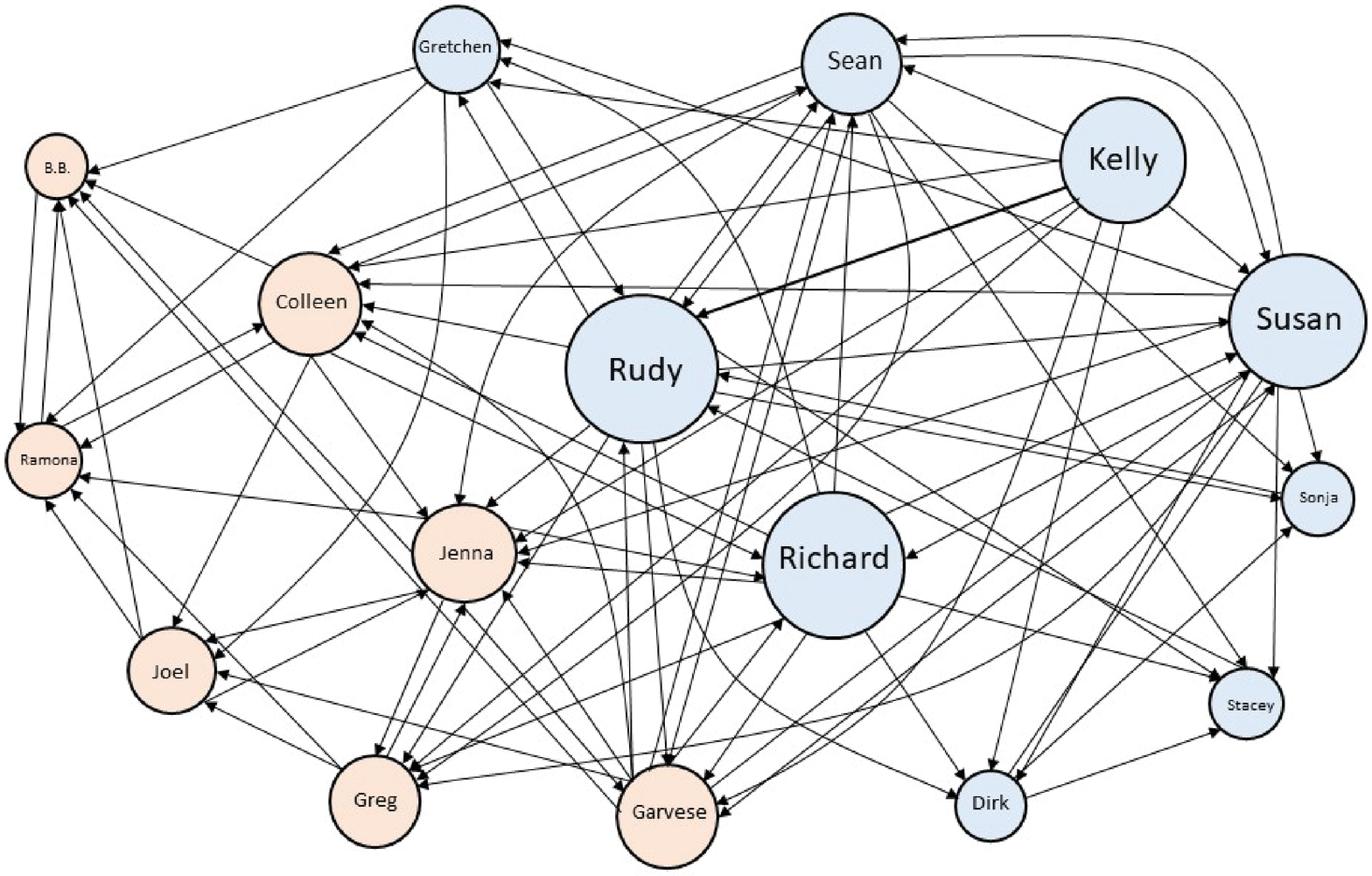,width=5in,height=3in}
\end{figure}
\end{center}

Note that Richard, the Sole Survivor of the season, has one of the highest closeness and CON scores. Rudy and Susan have higher scores, however. We note that Kelly won individual immunity several
times near the end of the game, and her voting out Rudy and Susan was a deciding factor in Richard's win.  We also note that comparing betweenness of players is inconclusive as a predictor of leaders. For example, we computed Richard's betweenness as 28.7, Kelly's as 0, and Rudy's as 36.5. One explanation of this is that leaders tend to have lower in-degree, which may reduce the number of
paths traversing through them. As such, we do not include betweenness scores for other seasons.

\subsection{China}

We next turn to Survivor: China, which was chosen because it represents a sample after the game was better known, and contestants better understood which strategies to employ in the game.

\begin{center}
\begin{tabular}{|l|l|l|l|c|l|}
\hline

Name & Tribe & ID & OD & C & CON \\ \hline \hline

Todd & Fei Long & 5 & 9 & 0.765 & 49 \\ \hline
Courtney & Fei Long & 0 & 9 & 0.667 & 39 \\ \hline
Amanda & Fei Long & 0 & 9 & 0.737 & 49 \\ \hline
Denise & Fei Long & 3 & 9 & 0.722 & 40 \\ \hline
Peih-Gee & Zhan Hu & 8 & 10 & 0.722 & 41 \\ \hline
Erik & Zhan Hu & 5 & 9 & 0.722 & 41 \\ \hline
James & Fei Long & 9 & 6 & 0.591 & 31 \\ \hline
Frosti & Zhan Hu & 7 & 7 & 0.65 & 39\\ \hline
Jean-Robert & Fei Long & 12 & 4 & 0.5 & 23 \\ \hline
Jaime & Zhan Hu & 7 & 5 & 0.481 & 26 \\ \hline
Sherea & Zhan Hu & 6 & 4 & 0.448 & 24 \\ \hline
Aaron & Fei Long & 3 & 2 & 0.406 & 12\\ \hline
Dave & Zhan Hu & 6 & 3 & 0.382 & 11 \\ \hline
Leslie & Fei Long & 6 & 1 & 0.342 & 9 \\ \hline
Ashley & Zhan Hu & 8 & 2 & 0.464 & 10\\ \hline
Chicken & Zhan Hu & 5 & 1 & 0.333 & 6 \\ \hline
\end{tabular}
\begin{figure}[h!]
\epsfig{figure=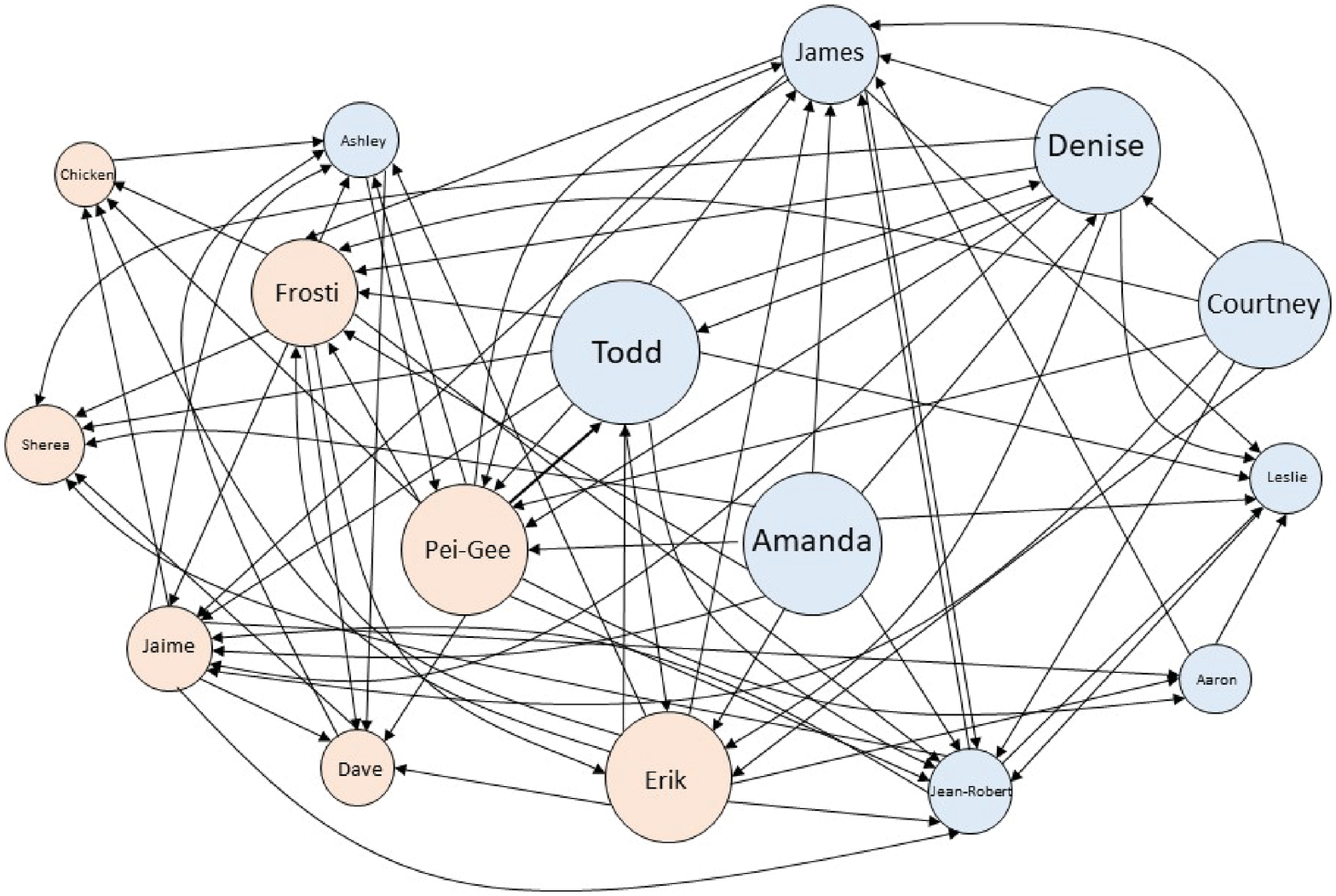,width=5in,height=3.5in}
\end{figure}
\end{center}
In this season, it is evident that Todd, the Sole Survivor, is the clear front-runner for Sole Survivor based on his high closeness and CON scores. Courtney and Amanda emerge also as leaders based on
their scores.

\subsection{Game Changers}

We next analyzed Survivor: Game Changers, as the second-to-last season of the show.
\begin{center}
\begin{tabular}{|l|l|l|l|c|l|} \hline

Name & Tribe & ID & OD & C & CON \\ \hline \hline

Sarah & Nuku & 3 & 13 & 0.692 & 64 \\ \hline
Brad & Nuku & 2 & 12 & 0.643 & 49\\ \hline
Troyzan & Mana & 2 & 12 & 0.643 & 55 \\ \hline
Tai & Nuku & 12 & 13 & 0.72 & 56 \\ \hline
Aubry & Mana & 9 & 13 & 0.72 & 61 \\ \hline
Cirie & Nuku & 0 & 8 & 0.613 & 45 \\ \hline
Michaela & Mana & 11 & 11 & 0.643 & 51 \\ \hline
Andrea & Nuku & 14 & 8 & 0.581 & 39 \\ \hline
Sierra & Nuku & 15 & 7 & 0.581 & 34 \\ \hline
Zeke & Nuku & 11 & 6 & 0.6 & 39 \\ \hline
Debbie & Nuku & 6 & 7 & 0.545 & 32 \\ \hline
Ozzy & Nuku & 7 & 4 & 0.5 & 22\\ \hline
Hali & Mana & 8 & 5 & 0.474 & 28 \\ \hline
Jeff & Mana & 6 & 5 & 0.529 & 33 \\ \hline
Sandra & Mana & 5 & 5 & 0.581 & 34 \\ \hline
JT & Nuku & 3 & 2 & 0.45 & 18 \\ \hline
Malcom & Mana & 5 & 3 & 0.439 & 24 \\ \hline
Caleb & Mana & 5 & 3 & 0.4 & 21 \\ \hline
Tony & Mana & 7 & 2 & 0.439 & 15 \\ \hline
Ciera & Mana & 9 & 1 & 0.4 & 8 \\ \hline
\end{tabular}
\begin{figure}[h!]
\epsfig{figure=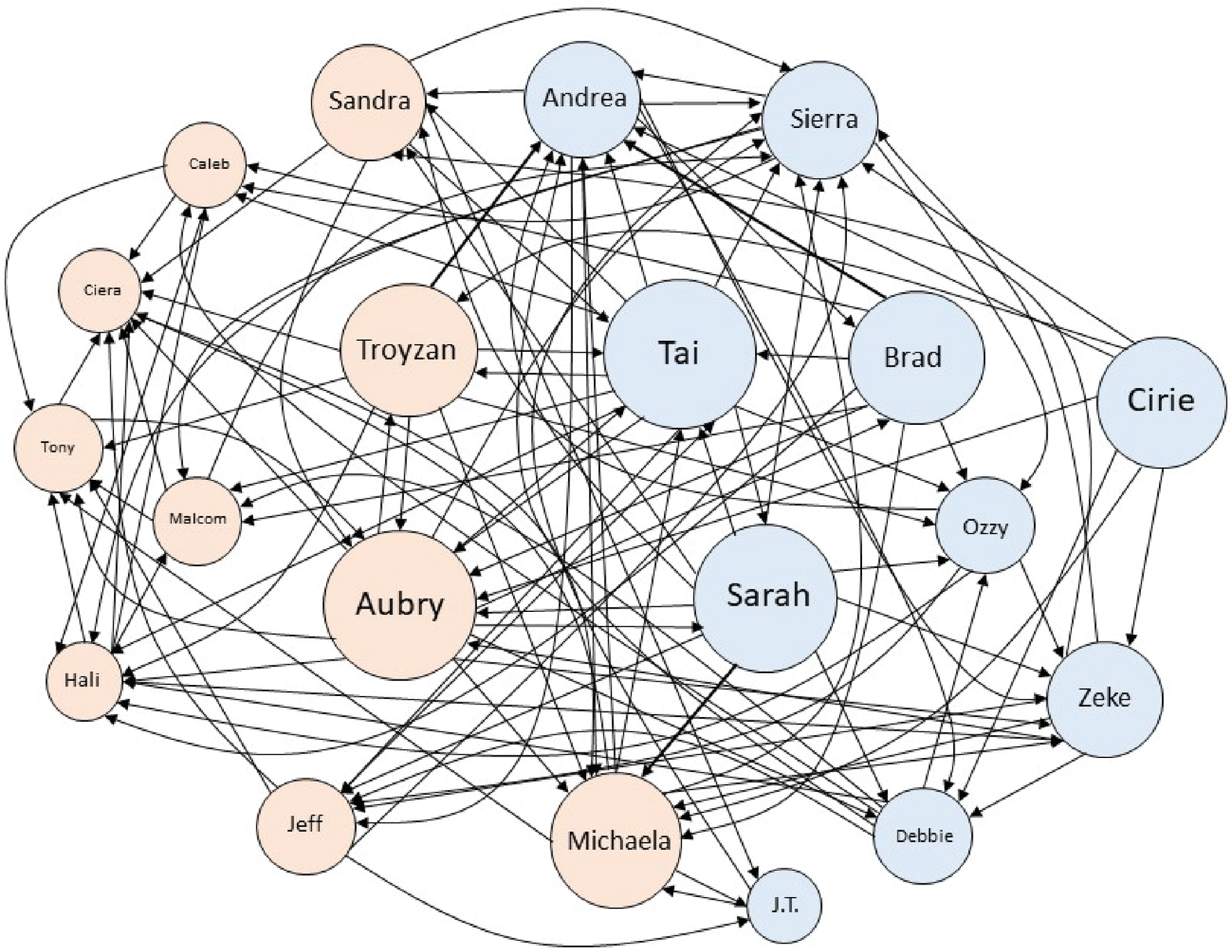,width=4in,height=3in}
\end{figure}
\end{center}

In this season, the Sole Survivor Sarah has high closeness and CON scores, but Tai and Aubry have higher closeness scores. Note, however, both players have high in-degrees which likely disadvantaged them.

\subsection{HHH}

We now turn to the most recent season of Survivor, Survivor: Heroes vs Healers vs Hustlers (or HHH, for short). The following table contains network data for Survivor: HHH.
\begin{center}
\begin{tabular}{|l|l|l|l|c|l|}
\hline

Name & Tribe & ID & OD & C & CON \\ \hline \hline
Ben & Levu & 11 & 11 & 0.63 & 41 \\ \hline
Chrissy & Levu & 7 & 13 & 0.68 & 44 \\ \hline
Ryan & Yawa & 2 & 14 & 0.708 & 47 \\ \hline
Devon & Yawa & 2 & 11 & 0.708 & 55 \\ \hline
Mike & Soko & 9 & 9 & 0.63 & 37 \\ \hline
Ashley & Levu & 8 & 10 & 0.607 & 46 \\ \hline
Lauren & Yawa & 3 & 7 & 0.63  & 39\\ \hline
Joe & Soko & 12 & 6 & 0.607 & 26\\ \hline
JP & Levu & 6 & 8 & 0.586 & 25\\ \hline
Cole & Soko & 7 & 4 & 0.531 & 26 \\ \hline
Desi & Soko & 11 & 3 & 0.515 & 9 \\ \hline
Jessica & Soko & 7 & 1 & 0.415 & 6 \\ \hline
Ali & Yawa & 3 & 4 & 0.5 & 19 \\ \hline
Roark & Soko & 3 & 1 & 0.415 & 6 \\ \hline
Alan & Levu & 2 & 2 & 0.415 & 11 \\ \hline
Patrick & Yawa & 5 & 2 & 0.405 & 6 \\ \hline
Simone & Yawa & 5 & 1 & 0.293 & 4 \\ \hline
Katrina & Levu & 5 & 1 & 0.386 & 5 \\ \hline
\end{tabular}
\begin{figure}[h!]
\epsfig{figure=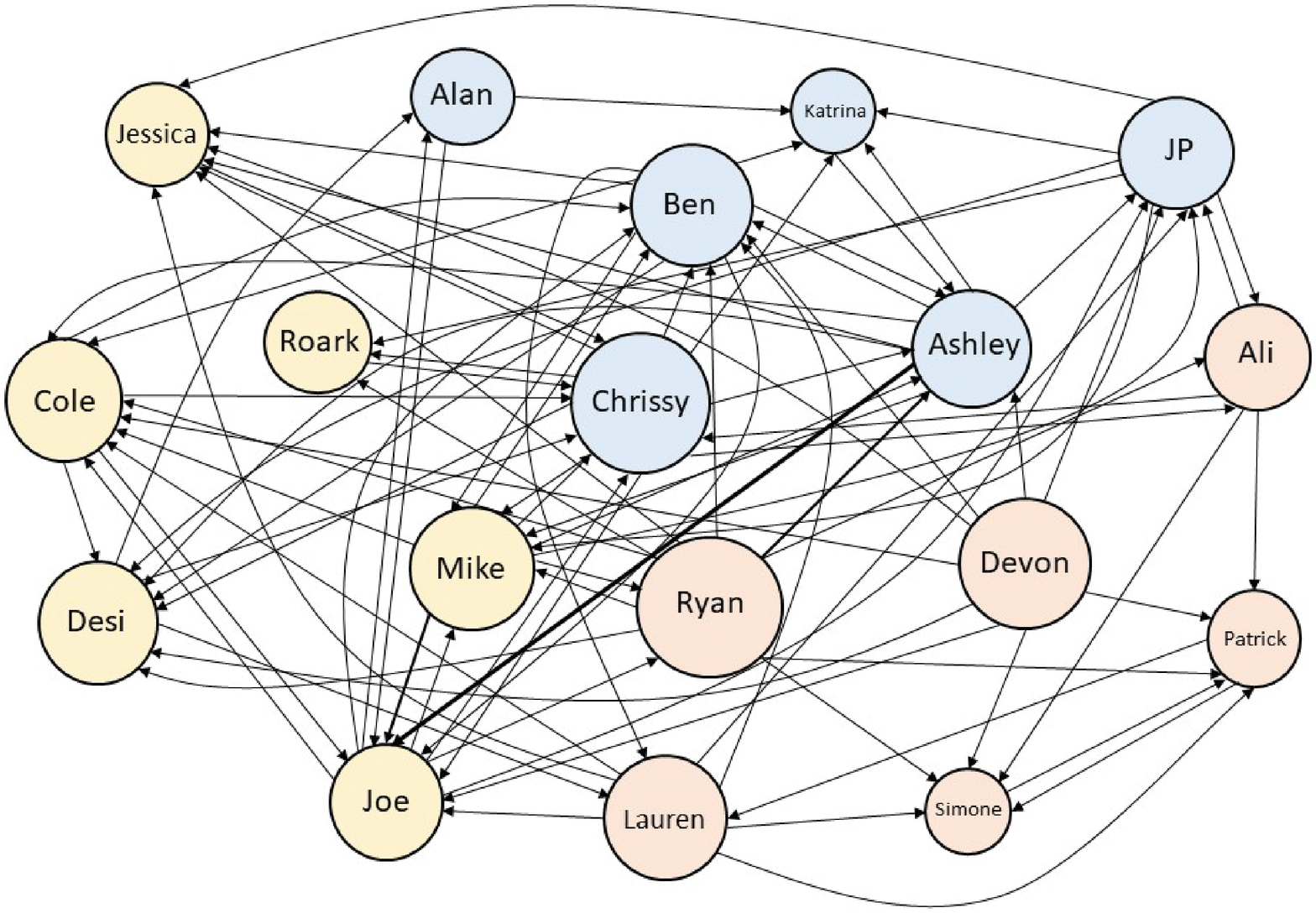,width=5in,height=3.5in}
\end{figure}
\end{center}

The finalists of this season were Ben, Chrissy and Ryan. Ryan and Devon had the highest overall closeness and highest overall CON scores, followed by Chrissy. However, Ben, the Sole Survivor, had lower scores than the other finalists; he secured his place in the final three by playing the hidden immunity idol three times.

\subsection{Big Brother}

Given the success of the DCH in Survivor, we turned to data from another social game Big Brother, focusing on Season 12.
\begin{center}
\begin{tabular}{|l|l|l|l|c|l|}
\hline
Name & ID & OD & C & CON \\ \hline \hline
Hayden & 3 & 16 & 0.923 & 44 \\ \hline
Lane & 3 & 10 & 0.857 & 46 \\ \hline
Enzo & 4 & 9 & 0.8 & 48 \\ \hline
Britney & 4 & 10 & 0.8 & 43 \\ \hline
Regan & 5 & 8 & 0.706 & 49 \\ \hline
Brendon & 7 & 9 & 0.706 & 40 \\ \hline
Matt & 9 & 7 & 0.632 & 35 \\ \hline
Kathy & 7 & 4 & 0.6 & 20 \\ \hline
Rachel & 8 & 6 & 0.667 & 24\\ \hline
Kristen & 7 & 3 & 1 & 25 \\ \hline
Andrew & 9 & 2 & 1 & 17 \\ \hline
Monet & 8 & 1 & 1 & 10 \\ \hline
Annie & 11 & 0 & 0 & 0 \\ \hline
\end{tabular}
\end{center}
\begin{figure}[h!]
\epsfig{figure=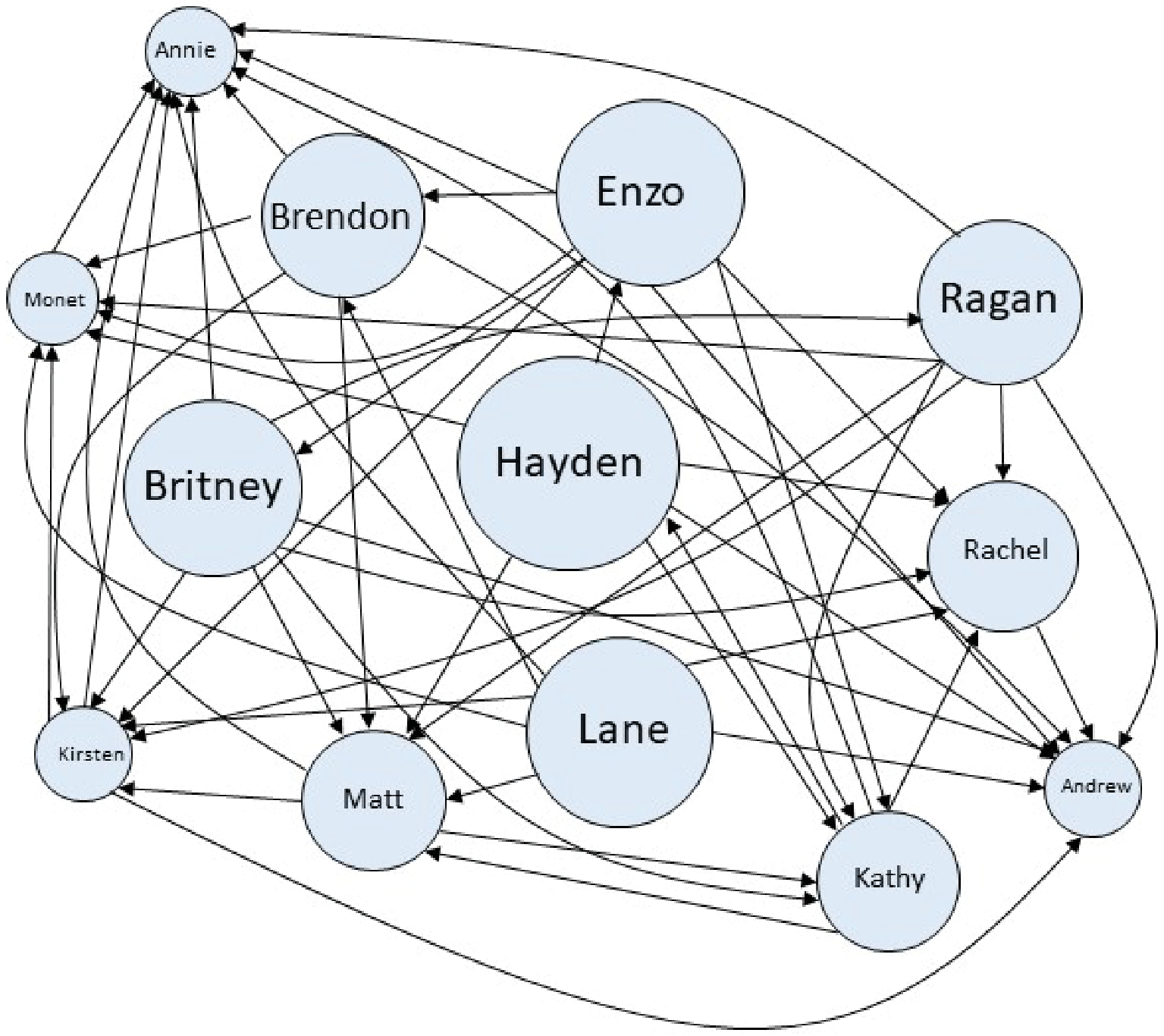,width=4in,height=3.5in}
\end{figure}

Hayden, the winner of the season, is the clear frontrunner with regards to closeness and CON scores, with HouseGuests Lane and Enzo rounding out the top three.

\subsection{CON scores}

There are 35 seasons in Survivor and 20 for Big Brother. In this section, we summarize that data. We are interested in knowing if a high CON score correlates with being the winner. To test this, we check whether the winner of a particular season has a CON score within the top three or five CON scores out of every player from that season. As displayed in the table below, 68.6$\%$ of winners in Survivor had a top three CON score, and 94.3$\%$ of them have a top five CON score.

We compare the CON score to two other well-known rankings: PageRank and Jaccard Similarity scores. Jaccard Similarity is a type of normalized CON score, and both of these methods are commonly used in ranking; see, for example, \cite{Gleich-2015-prbeyond} and~\cite{Similarity}. Note that we computed PageRank scores on the \emph{reverse} of the network discussed in Section~\ref{model}. The table shows that the CON scores are the best predictor for winners in Survivor, while PageRank is a slightly better predictor in Big Brother. Furthermore, we calculate the probability of the winner appearing in a random set of three or five, under the \emph{random set} column. This probability varies depending on the size of the network (that is, the number of players). We see, for example, that the probability of a winner being in a random set of three in Survivor is between 15$\%$ and $18.8\%$. In all cases, these probabilities are lower than the CON scores, which suggests that the result of the winner having one of the largest CON scores is not due to random chance.

\begin{center}
	\begin{tabular}[H]{ m{2.7cm}  m{1.5cm} m{1.5cm}  m{1.5cm}  m{2.3cm}  m{2cm} }
		      & & \textbf{CON} & \textbf{Page Rank} & \textbf{Jacard Similarity} & \textbf{random set} \\ \hline
		     \textbf{Survivor} & {\small\textbf{Top 3}}      & 68.6 & 54.3  & 54.3 & 15.0-18.8 \\
		     							&{\small\textbf{Top 5}}      & 94.3  & 88.6 & 80.0 & 25.0-31.3 \\
		     \textbf{Big Brother} & {\small\textbf{Top 3}} & 60.0  & 80.0 & 25.0 & 17.6-30.0 \\
		     							& {\small\textbf{Top 5}}     & 70.0  & 100   & 55.0 & 29.4-50.0\\
	\end{tabular}
\end{center}

\subsection{Alliances}

In addition to predicting winners, we analyzed alliances in the various seasons and computed their edge density. All the alliances conform to the DCH as they form near independent sets. Some
alliances have relatively high edge density, as we note in the Tagi alliance in Borneo (which includes the sole survivor Richard). Nevertheless, narrowing down the alliances to subsets of finalists
appears to reduce the edge density. For example, in the Tagi alliance, the edge density of the subsets $\{$Kelly,Richard$\}$ is 1/2 and $\{$Richard,Rudy$\}$ is 0. Analogously, in the Fie-Long
alliance in Survivor: China, the subset $\{$Amanda, Courtney, Todd$\}$ has edge density 0. 


\begin{center}
\begin{tabular}{| m{3cm} | m{1.6cm} | m{1.75cm} | m{7cm} | l |}
	\hline
	\textbf{Season} & \textbf{Winner} & \textbf{Finalists} & \textbf{Alliances} & \textbf{ED} \\ \hline
	\multirow{2}{*}{\parbox{1.75cm}{Borneo}}&\multirow{2}{*}{Richard}& \multirow{2}{*}{Kelly}& \textit{Barbecue}: Colleen, Jenna, Gervase&1.667\\
	& & & \textit{Tagi}: Richard, Rudy, Susan, Kelly&1.5\\
	\hline
	\multirow{2}{*}{\parbox{1.75cm}{China}}&\multirow{2}{*}{Todd}&Courtney& \textit{Fei Long}: Todd, Courtney, Amanda, Aaron, Denise, James, Frosti&0.667\\
	& &Amanda& \textit{Zhan Hu}: Peih-Gee, Erik, Jaime&0.0\\
	\hline
	\multirow{2}{*}{\parbox{1.75cm}{Game Changers}}&\multirow{2}{*}{Sarah}&Brad& \textit{Power Six}: Sarah, Brad, Troyzan, Sierra, Debbie, Tai&0.933\\
	& &Troyzan& \textit{Tavua}: Aubry, Cirie, Michaela, Ozzy, Andrea, Zeke, Sarah&1.238\\
	\hline
	\multirow{3}{*}{\parbox{2.15cm}{HHH}}&\multirow{3}{*}{Ben}&Chrissy& \textit{Healers}: Joe, Desi, Jessica, Cole, Mike&0.6\\
	& &Ryan& \textit{The Round Table}: Chrissy, Ryan, Devon, JP, Ben, Ashley, Lauren&0.905\\
	& & & \textit{Final Four}: Ashley, Lauren, Ben, Devon&1.333\\
	\hline
	\multirow{1}{*}{\parbox{2.75cm}{Big Brother 12}}&\multirow{1}{*}{Hayden}&Lane& \textit{The Brigade}: Enzo, Hayden, Lane, Matt&0.5\\
	\hline
\end{tabular}
\end{center}

We list the edge densities for each alliance in the Appendix, along with the edge density for the entire graph. There may be some use in exploring to what extent alliances have smaller edge density than that of the entire graph. As already discussed, the edge density of an alliance can become much lower when removing players who play against their alliance. That being said, 60$\%$ of the Survivor seasons have an alliance with a lower edge density than the edge density for the total graph, and 95$\%$ of Big Brother seasons have an alliance with a lower edge density than the edge density for the total graph. More exploration is needed to understand the relationship between the edge densities of alliances and leaders.

\section{Discussion and future work}\label{conc}

We introduced the notion of dynamic competition networks and studied their properties. The Dynamic Competition Hypothesis (DCH) was presented, which resolves dynamic competition networks arising from social networks into
alliances, detects leaders, and measures the relative strength of alliances. The DCH was tested with voting data from all seasons of the U.S. television social game shows Survivor and Big Brother. In all seasons
and as predicted by the DCH, alliances correspond to near independent sets, CON scores accurately determine leaders of alliances, and leaders are detected via their CON scores and closeness.

In future work, we will mine data from all international seasons of Survivor and Big Brother (our current analysis uses only seasons from a single country). We will also look for other data sets to further validate the DCH more broadly, within the
lens of structural balance theory and social network analysis. A weakness of our current theory is that longer lasting members of a season accumulate more influence simply due to their survival. In particular, players in Survivor and Big Brother that survive longer in the game have a greater opportunity to improve their CON-scores and other metrics. In future work, we will therefore, evaluate data at earlier stages of the formation of the network. Other areas where we can explore the DCH are food webs, signed networks (by extracting the subgraph with negative signs), and geo-political networks. It would be interesting to invert the DCH to determine low ranked members of dynamic competition networks. Further, it would be useful
to develop a mathematical model predicting the evolution of dynamic competition networks, which provably simulates properties predicted by the DCH.

\end{document}